\documentclass[11pt]{article}
\usepackage[T1]{fontenc}
\usepackage[hyperref]{acl}
\usepackage{float}
\usepackage{times}
\usepackage{latexsym}
\usepackage{microtype}
\usepackage{inconsolata}
\usepackage{graphicx}
\usepackage{booktabs}
\usepackage{multirow}
\usepackage{amsmath}
\usepackage{listings}
\usepackage{xcolor}
\usepackage{url}
\usepackage{array}

\definecolor{codegreen}{rgb}{0.25,0.49,0.48}
\definecolor{codegray}{rgb}{0.5,0.5,0.5}
\definecolor{codepurple}{rgb}{0.58,0,0.82}
\definecolor{backcolour}{rgb}{0.97,0.97,0.97}

\lstdefinelanguage{Rust}{
  keywords={fn, let, if, else, match, struct, impl, pub, use, mod, return,
            async, await, move, tokio, self, true, false, None, Some, Ok, Err},
  keywordstyle=\color{codepurple}\bfseries,
  comment=[l]{//},
  commentstyle=\color{codegreen}\itshape,
  stringstyle=\color{orange},
  morestring=[b]",
  sensitive=true
}

\title{Application-Layer Dual Memory for Conversational AI:\\
       Achieving Virtually Unbounded Context Without Model Modification}

\author{
\footnotesize
\begin{tabular}{@{}ccc@{}}
\textbf{Rajendra Narayan Jena} & \textbf{Rajan Padmanabhan} & \textbf{Sankar Arumugam} \\
Infosys Limited & Infosys Limited & Infosys Limited \\
\texttt{rajendra\_jena@infosys.com} & \texttt{Rajan\_P@infosys.com} & \texttt{sankaran\_arumugam@infosys.com}
\end{tabular}
}

\begin{document}
\maketitle

\begin{abstract}
Large language models (LLMs) operate within fixed context windows that fundamentally
limit conversational continuity. When context fills, compaction discards history
irreversibly; when sessions end, all memory resets to zero. Existing solutions---larger
context windows, retrieval-augmented generation for knowledge bases, and
memory-augmented architectures such as MemGPT---either require model modification,
impose provider lock-in, or do not address the compaction continuity problem.
We present \textbf{CALMem (Conversational Application-Layer Memory)}, an application-layer
dual memory architecture that gives LLM-based conversational assistants virtually
unbounded effective context without any modification to the underlying model.
CALMem combines two complementary memory subsystems: an \emph{episodic memory} layer
built on sliding-window vector embeddings of conversation history, and a
\emph{semantic memory} layer of agent-writable structured facts.
A token-budget-adaptive injection mechanism, called the
\textbf{MOIM (Message of Injected Memory)}, automatically retrieves and injects relevant
past context each turn, scaling injection depth inversely with context pressure.
A key contribution is \textbf{intra-session retrieval}: compacted-away turns from the
\emph{current} session remain searchable, closing a gap unaddressed by prior work.
The system is implemented as a pure application layer in a production Rust codebase,
is provider-agnostic, and degrades to original LLM behaviour with zero overhead when
disabled. We describe the architecture, design decisions, and performance
characteristics, and analyse the trade-offs that guided each implementation choice.
\end{abstract}

\section{Introduction}

The context window is the fundamental unit of memory for a large language model.
Everything the model ``knows'' during a conversation must fit within it.

Current frontier models offer windows that have grown dramatically: GPT-5.2 provides
400,000 tokens \cite{OPENAI2026}, Claude~4.6 Sonnet extends to 1,000,000
\cite{ANTHROPIC2026}, Gemini~2.0 Pro reaches 2,000,000 \cite{GOOGLE2026}, and
Alibaba's Qwen3-30B supports up to 1,000,000 tokens \cite{ALIBABA2026}. Despite these
remarkable capacities, two structural problems remain unsolved. First, windows are still
finite: long-running, task-intensive conversations---particularly in software engineering,
research, or multi-day projects---exhaust even million-token windows. Second, and more
fundamentally, every new conversation begins at zero regardless of window size.
A 2,000,000-token window provides no memory of yesterday's session.

Practical deployments handle this through \emph{compaction}: when the context approaches
capacity, older turns are summarised and discarded from the active window. Compaction
preserves the overall thread of conversation but loses detail. Once a turn is compacted
away, any specific fact, decision, or piece of work it contained becomes inaccessible to
the model---permanently, unless re-introduced through user effort.

The cross-session problem is worse. Every new conversation begins from a blank slate.
A user who spent three sessions last week debugging a complex system must re-explain
the entire context at the start of session four. The model has no memory of any of it.

These are not problems with model capability. They are structural consequences of the
context window as an architectural primitive. The question is not whether to build larger
windows---hardware and attention complexity impose real limits \cite{VASWANI2017}---but
whether the application layer can compensate intelligently.

There is also a third, practical problem: \textbf{cost}. LLM providers bill per token
in context. Sending a 2,000,000-token window to the model on every turn costs
proportionally more than sending a focused, compact context. For an application that
manages memory intelligently---retrieving only the $\sim$750 tokens of relevant history
each turn rather than retaining millions of tokens in the active window---the per-turn
API cost can be one to two orders of magnitude lower than the na\"{i}ve ``use the
largest window available'' approach. CALMem is not only a memory architecture; it is
inherently a cost-efficiency architecture.

\textbf{Retrieval-Augmented Generation} (RAG) \cite{LEWIS2020} addresses this for
static knowledge bases: pre-index a corpus, retrieve relevant passages at query time,
inject into the prompt. But applying RAG to conversation history introduces problems not
present in document retrieval. Conversation is not a corpus: it is a temporally ordered,
contextually entangled stream.

\textbf{MemGPT} \cite{PACKER2023} introduced the idea of OS-inspired virtual context
for LLMs, treating the context window as RAM and external storage as disk. This is a
powerful framing, but it requires custom prompting infrastructure, relies on the model
itself to make memory management decisions, and does not distinguish episodic from
semantic memory.

We propose a different approach: build the memory system entirely in the application
layer, invisible to the model, requiring no prompt changes, no fine-tuning, and no
modification to the underlying LLM provider.

This paper makes the following contributions:
\begin{enumerate}
  \item \textbf{CALMem}, a dual memory architecture combining episodic retrieval (vector
        embeddings of conversation history) and semantic retrieval (structured
        agent-writable facts), deployed as a pure application layer.
  \item \textbf{Token-budget-adaptive injection}: a MOIM mechanism that scales retrieval
        depth proportionally to available context headroom, preventing memory injection
        from accelerating the compaction it is designed to mitigate.
  \item \textbf{Intra-session continuity}: episodic retrieval operates over
        compacted-away turns from the current session, not just prior sessions---
        recovering context that no existing system retrieves.
  \item \textbf{Zero-regression design}: the system degrades to standard LLM behaviour
        when disabled, with no latency overhead, no schema dependency, and no prompt
        structure changes.
  \item A concrete implementation in a production Rust system with analysis of design
        decisions, performance characteristics, and observed trade-offs.
\end{enumerate}

The rest of the paper is organised as follows. Section~\ref{sec:related} reviews related
work. Section~\ref{sec:arch} describes the overall architecture. Sections~\ref{sec:episodic}
and~\ref{sec:semantic} detail the two memory subsystems. Section~\ref{sec:moim} covers
token-budget-adaptive injection. Section~\ref{sec:perf} discusses performance
optimisation. Section~\ref{sec:impl} describes the implementation.
Section~\ref{sec:eval} presents evaluation. Section~\ref{sec:discussion} discusses
limitations and future work. Section~\ref{sec:conclusion} concludes.

\section{Related Work}
\label{sec:related}

\subsection{Retrieval-Augmented Generation}

The original RAG framework \cite{LEWIS2020} augments autoregressive language models with
a non-parametric retrieval component: a dense passage retrieval (DPR)
\cite{KARPUKHIN2020} model over a fixed document corpus. RAG was designed for
open-domain question answering over static knowledge bases, not for conversational
memory. Subsequent work has extended RAG to dialogue settings
\cite{XU2022,ZHONG2022}, but these approaches typically treat conversation history as a
flat document collection, losing the sequential and session structure that makes
conversational memory different from document retrieval.

\subsection{Long-Context Language Models}

A parallel line of work extends the context window itself through architectural
innovations: sparse attention \cite{CHILD2019}, linear attention
\cite{KATHAROPOULOS2020}, and sliding window attention \cite{BELTAGY2020}. More
recently, FlashAttention \cite{DAO2022} makes very large windows computationally
tractable. Despite these advances, the context window remains finite. More importantly,
larger windows do not address the cross-session problem: each new conversation still
begins with zero memory regardless of window size.

\subsection{Memory-Augmented Language Models}

MemGPT \cite{PACKER2023} introduced a compelling operating-system analogy: the context
window as RAM, with a hierarchical external memory as disk, managed by the model itself
through explicit function calls. This framework has inspired a family of agent memory
architectures \cite{SUMERS2023,WANG2023}. The key limitation is that memory management
is delegated to the model.

Generative Agents \cite{PARK2023} demonstrated long-horizon memory for simulated social
agents by combining reflection, synthesis, and retrieval. The system is powerful but
purpose-built for simulation; it is not designed as a drop-in layer for general-purpose
conversational assistants.

\subsection{Conversation Summarisation}

An alternative to retrieval is summarisation: periodically compress conversation history
into a running summary \cite{WU2021,CHEN2023}. Summarisation loses detail by design.
A retrieval-based approach recovers specific detail on demand rather than trading it for
compression. Our system is not an alternative to summarisation but a complement:
compaction runs when necessary, and the retrieval layer recovers relevant compacted
content.

\subsection{Recent RAG and Agent Memory Advances (2024--2026)}

RAPTOR \cite{SARTHI2024} introduces recursive abstractive processing: it builds a tree
of progressively summarised abstractions using an LLM, enabling retrieval at multiple
levels of granularity. CALMem deliberately avoids recursive abstraction for conversation
history---raw chunk preservation maintains verbatim detail that summarisation would
erase, and abstraction requires an additional LLM call per indexed message.

HippoRAG \cite{GUTIERREZ2024} draws inspiration from hippocampal memory indexing theory,
using a knowledge graph to link named entities and Personalised PageRank for retrieval
scoring. The biological framing parallels CALMem's own grounding in Tulving's
episodic/semantic memory distinction \cite{TULVING1972}. However, HippoRAG targets
static knowledge bases; its graph construction is not suited to the append-only,
session-structured nature of conversation history.

GraphRAG \cite{EDGE2024} applies community detection on entity-relationship graphs to
support global queries over large corpora. This is powerful for corpus-level
summarisation but architecturally misaligned with personal conversational memory.

The period 2024--2026 has seen MemGPT evolve into Letta, a production-grade agentic
memory platform, and the emergence of dedicated memory-layer services (Mem0, Zep).
These systems validate the demand for application-layer memory but typically operate as
external services, introducing network latency, data privacy concerns, and provider
dependency. CALMem runs entirely in-process with zero network dependency.

\subsection{Positioning}

CALMem occupies a position that prior work does not: it is (a) a pure application layer
requiring no model modification, (b) aware of both compaction state and session
boundaries, (c) equipped with two complementary memory types, and (d)
token-budget-adaptive. Table~\ref{tab:comparison} summarises the comparison.

\begin{table*}[t]
\centering
\caption{Comparison with related approaches.}
\label{tab:comparison}
\resizebox{\linewidth}{!}{%
\begin{tabular}{llllll}
\toprule
\textbf{System} & \textbf{Layer} & \textbf{Memory Types} & \textbf{Budget-Aware} & \textbf{Intra-Session} & \textbf{Provider-Agnostic} \\
\midrule
RAG \cite{LEWIS2020}                  & Application   & Episodic only            & No  & No  & Yes     \\
MemGPT \cite{PACKER2023}              & Prompt/Agent  & Hierarchical             & No  & No  & Partial \\
Gen.\ Agents \cite{PARK2023}          & Application   & Episodic + Reflection    & No  & No  & Yes     \\
RAPTOR \cite{SARTHI2024}              & Application   & Hierarchical (tree)      & No  & No  & Yes     \\
HippoRAG \cite{GUTIERREZ2024}         & Application   & Graph-indexed episodic   & No  & No  & Yes     \\
Long-context LLMs \cite{GOOGLE2026}   & Model         & Up to 2M tokens          & N/A & Yes & No      \\
\textbf{CALMem (this work)}           & Application   & Episodic + Semantic      & \textbf{Yes} & \textbf{Yes} & \textbf{Yes} \\
\bottomrule
\end{tabular}}
\end{table*}

\section{System Architecture}
\label{sec:arch}

\subsection{Overview}

\begin{figure}[H]
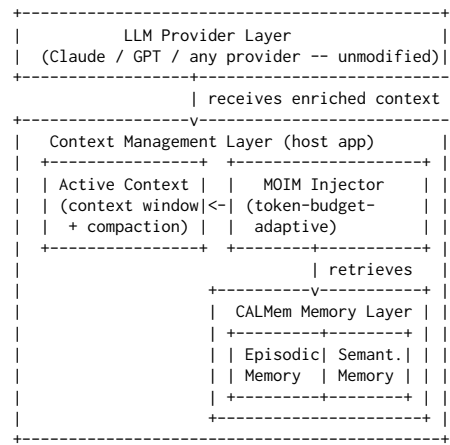

\centering
{\scriptsize
\begin{verbatim}
+---------------------------------------------+
|           LLM Provider Layer                |
|  (Claude / GPT / any provider -- unmodified)|
+------------------+---------------------------+
                   | receives enriched context
+------------------v---------------------------+
|   Context Management Layer (host app)       |
|  +----------------+  +--------------------+ |
|  | Active Context |  |   MOIM Injector    | |
|  | (context window|<-| (token-budget-     | |
|  |  + compaction) |  |  adaptive)         | |
|  +----------------+  +--------+-----------+ |
|                               | retrieves   |
|                    +----------v-----------+ |
|                    |  CALMem Memory Layer | |
|                    | +---------+--------+ | |
|                    | | Episodic| Semant.| | |
|                    | | Memory  | Memory | | |
|                    | +---------+--------+ | |
|                    +----------------------+ |
+---------------------------------------------+
\end{verbatim}
}
\caption{CALMem three-layer architecture. The LLM provider receives an enriched context
assembled by the MOIM injector from both active context and memory layer retrieval.}
\label{fig:arch}
\end{figure}

CALMem is structured as three cooperating layers, illustrated in Figure~\ref{fig:arch}.

\subsection{The Three Layers}

\textbf{LLM Provider Layer.} The underlying language model, accessed via any standard
API. CALMem makes no assumptions about the model beyond the existence of a
message-based conversation interface. No fine-tuning, no special tokens, no prompt
structure changes are required.

\textbf{Context Management Layer.} Manages the active conversation window and handles
compaction when the window approaches capacity. CALMem integrates non-invasively: a hook
in the message persistence path triggers background indexing; the MOIM injector runs
once per turn to prepend retrieved memory to the system context.

\textbf{Memory Layer.} The novel contribution. Two independent stores with complementary
retrieval characteristics, described in Sections~\ref{sec:episodic}
and~\ref{sec:semantic}.

\subsection{The MOIM Interface}

The \textbf{MOIM (Message of Injected Memory)} is the interface between the memory layer
and the context management layer. It is a structured text block prepended to the system
prompt each turn, containing the most relevant retrieved content from both memory
subsystems. The model receives this as context---it does not call retrieval explicitly
(unlike MemGPT). The content and size of the MOIM block adapt to available context
headroom, as described in Section~\ref{sec:moim}.

\subsection{Indexing Pipeline}

Every message persisted to the conversation database triggers a background indexing task
(Figure~\ref{fig:pipeline}). This is non-blocking: the message commit completes before
indexing begins. The critical invariant is that the conversation database is always the
source of truth; the memory index is a derived, recoverable structure.

\begin{figure}[H]
\centering
{\footnotesize
\begin{verbatim}
add_message()
      |
      v
  tx.commit()
  [message durable in SQLite]
      |
      v  [tokio::spawn]
  +-- background task -------+
  | extract_chunks(msg)      |
  | embed() via fastembed    |
  | INSERT OR IGNORE         |
  |   rag_chunks             |
  | update cache (if warm)   |
  +--------------------------+
\end{verbatim}
}
\caption{Indexing pipeline. The message is committed to the database before any indexing
work begins, ensuring durability regardless of indexing outcome.}
\label{fig:pipeline}
\end{figure}

The following Rust excerpt shows the non-blocking dispatch at the message persistence
call site.

\begin{lstlisting}[language=Rust, caption={Non-blocking indexing dispatch after message commit.}]
// Message is durable in SQLite before indexing starts.
tx.commit().await?;

if is_rag_enabled() {
    let (sid, mid, text) = (
        session_id.to_string(),
        msg.id.clone(),
        msg.content.clone()
    );
    tokio::spawn(async move {
        let indexer = RagIndexer::instance();
        if let Err(e) = indexer.index_message(
                &sid, &mid, &text).await {
            tracing::warn!("RAG indexing failed: {}", e);
        }
    });
}
\end{lstlisting}

\section{Episodic Memory: Vector-Based Conversational Retrieval}
\label{sec:episodic}

\subsection{Design Rationale}

Episodic memory captures the sequential, experiential record of conversation: what was
said, when, in what context. Retrieval from episodic memory is \emph{fuzzy}---the user
asking ``what was that approach we discussed for the authentication bug?'' does not know
the exact wording of the relevant turn.

Dense vector retrieval is the natural fit. We use \textbf{all-MiniLM-L6-v2} via the
fastembed library \cite{REIMERS2019}---a 384-dimensional sentence embedding model
implemented in pure Rust via ONNX, requiring no Python runtime or sidecar process. The
model produces $\sim$3\,ms embeddings on CPU, is $\sim$23\,MB, and achieves strong
performance on the MTEB semantic textual similarity benchmarks \cite{MUENNIGHOFF2022}
while remaining lightweight enough for a desktop application.

\subsection{Chunking Strategy}

We apply \textbf{sliding-window chunking} with overlap: each message is divided into
chunks of at most 1,000 characters ($\approx$250 tokens), with consecutive chunks
overlapping by 200 characters (Figure~\ref{fig:chunking}). The overlap ensures that
semantically related sentences straddling a chunk boundary are captured by at least one
chunk.

\begin{figure}[H]
\centering
{\footnotesize
\begin{verbatim}
Message text (2400 chars):
+---- Chunk 1 [0..1000] -----+
       +---- Chunk 2 [800..1800] ----+
              +--- Chunk 3 [1600..2400] ---+
       |<200>|            |<200>|
\end{verbatim}
}
\caption{Sliding-window chunking (size=1000, overlap=200, step=800).
Overlap preserves cross-boundary semantic context.}
\label{fig:chunking}
\end{figure}

Chunk boundaries are snapped to the nearest whitespace within the overlap window to
avoid mid-word cuts. All chunking operates on Unicode scalar values (characters), not
bytes, ensuring correctness for multi-byte UTF-8 text. Short messages ($\leq$1,000
characters) are stored as a single chunk---a fast path that avoids unnecessary
computation for the majority of conversational turns.

\subsection{Intra-Session Retrieval}

A critical and novel aspect of CALMem is that \textbf{intra-session retrieval} operates
over compacted-away turns from the \emph{current} session, not only past sessions.

Standard RAG over conversation history is typically applied cross-session. This misses
a significant class of relevant content: turns from the current session that have
already been compacted out of the active context window. ConvMem implements this by
filtering at search time: chunks from the current session whose message IDs appear in
the active context are excluded (they are already visible to the model); compacted-away
chunks are included.

\begin{lstlisting}[language=Rust, caption={Intra-session filter predicate applied during scoring.}]
// active_message_ids: HashSet of IDs in the context window.
// Compacted-away chunks pass through and are scored normally.
let exclude = chunk.session_id == current_session_id
    && active_message_ids.contains(&chunk.message_id);

if !exclude {
    let score = cosine_similarity(
        &query_vec, &chunk.embedding);
    if score >= SIMILARITY_THRESHOLD {
        results.push((score,
            chunk.session_id.clone(),
            chunk.chunk_text.clone()));
    }
}
\end{lstlisting}

This gives the model access to a seamless memory spanning: current active context
(directly visible) $\rightarrow$ compacted current session history (intra-session
retrieval) $\rightarrow$ prior sessions (cross-session retrieval).

\subsection{Retrieval Mechanics}

At search time, the query text is embedded using the same model as indexing. Cosine
similarity is computed against all stored embeddings. A minimum threshold of
\textbf{0.4} filters results not relevant enough to include. The top-$k$ results
(default $k=5$) are returned ranked by similarity.

The retrieval is exact (brute-force scan), not approximate. For the target deployment
scale (up to $\sim$100K chunks, representing months to years of daily use), exact
retrieval is computationally adequate and avoids the approximation error and cold-start
penalties of graph-based ANN indices such as HNSW \cite{MALKOV2018} (see
Section~\ref{sec:perf} for detailed analysis).

\section{Semantic Memory: Structured Agent-Writable Facts}
\label{sec:semantic}

\subsection{Design Rationale}

Episodic memory is a poor fit for precise, stable, explicitly-stated facts. A user who
tells the assistant their preferred programming language wants that fact retrieved with
100\% precision regardless of semantic distance. This motivates a second memory
subsystem: \textbf{semantic memory}, a structured key-value store that the agent writes
to explicitly and reads from automatically. Semantic memory in CALMem closely parallels
its cognitive science counterpart \cite{TULVING1972}: general, decontextualised
knowledge as opposed to episodic knowledge.

\subsection{Structure and Persistence}

Semantic memory is stored as a relational table (\texttt{memory\_facts}) with the
following schema:

\begin{lstlisting}[language=SQL, caption={Schema for the semantic memory store.}]
CREATE TABLE memory_facts (
    id         INTEGER PRIMARY KEY AUTOINCREMENT,
    session_id TEXT NOT NULL,
    category   TEXT NOT NULL DEFAULT 'general',
    key        TEXT NOT NULL,
    value      TEXT NOT NULL,
    created_at TIMESTAMP DEFAULT CURRENT_TIMESTAMP,
    FOREIGN KEY (session_id)
        REFERENCES sessions(id) ON DELETE CASCADE
);
CREATE UNIQUE INDEX ON memory_facts(session_id, key);
\end{lstlisting}

The \texttt{UNIQUE INDEX} on \texttt{(session\_id, key)} enables upsert semantics.
Categories (e.g., ``preference'', ``project'', ``configuration'') allow structured
filtering without schema changes.

\subsection{Agent Interface}

The agent interacts with semantic memory through four explicit tools exposed via the
Model Context Protocol (MCP):
\begin{itemize}
  \item \textbf{\texttt{remember\_fact}} --- write or update a fact: \texttt{(key, value, category)}
  \item \textbf{\texttt{recall\_facts}} --- search facts by key pattern and/or category, cross-session
  \item \textbf{\texttt{forget\_fact}} --- delete a fact by key within the current session
  \item \textbf{\texttt{search\_memory}} --- semantic search over episodic memory (bridges both subsystems)
\end{itemize}

\subsection{Automatic Injection vs Explicit Retrieval}

Semantic memory participates in MOIM injection automatically: all facts for the current
session are injected into every turn. Cross-session fact retrieval is available but not
injected automatically---the agent uses \texttt{recall\_facts} when it needs facts from
past sessions. This distinction prevents the injection block from growing unbounded as
the fact store accumulates across sessions.

\section{Token-Budget-Adaptive Injection}
\label{sec:moim}

\subsection{The Injection Problem}

Na\"{i}ve memory injection creates a paradox: if memory retrieval injects a fixed amount
of context each turn, and the context window is already filling with active conversation,
memory injection may consume capacity that accelerates compaction. The system designed to
extend context ends up shortening it.

\subsection{The Four-Tier Budget Model}

CALMem defines a context fill ratio $r = \text{current\_tokens} /
\text{context\_window\_capacity}$. The application-layer compaction threshold is set at
$r = 0.8$. MOIM injection uses four tiers keyed on this ratio (Figure~\ref{fig:tiers}):

\begin{figure}[H]
\centering
\footnotesize
\begin{tabular}{ll}
\toprule
\textbf{Fill ratio} & \textbf{MOIM budget} \\
\midrule
$r < 0.60$              & 5 chunks $\times$ 600 chars \\
$0.60 \leq r < 0.70$    & 3 chunks $\times$ 400 chars \\
$0.70 \leq r < 0.80$    & 2 chunks $\times$ 250 chars \\
$r \geq 0.80$           & Suppress injection \\
\bottomrule
\end{tabular}
\caption{MOIM token budget tiers. Injection depth scales inversely with context pressure.}
\label{fig:tiers}
\end{figure}

At $r \geq 0.80$, compaction is imminent; episodic injection is suppressed. Fact
injection (semantic memory) is not suppressed at any tier.

The tier selection maps cleanly to a single Rust match expression. A noteworthy
correctness detail: the boundaries are \textbf{literal constants} (\texttt{0.60},
\texttt{0.70}), not computed multiples of the compaction threshold. Computing
$0.8 \times 0.75$ in IEEE~754 \texttt{f32} arithmetic yields \texttt{0.6000000000000001},
which can misclassify values exactly at \texttt{0.60}.

\begin{lstlisting}[language=Rust, caption={MOIM budget tier selection using literal constants.}]
/// Returns (max_chunks, max_chars) or None to suppress.
fn moim_budget(r: f32) -> Option<(usize, usize)> {
    match r {
        r if r >= 0.80 => None,           // suppress
        r if r >= 0.70 => Some((2, 250)), // minimal
        r if r >= 0.60 => Some((3, 400)), // reduced
        _              => Some((5, 600)), // full
    }
}
\end{lstlisting}

\subsection{MOIM Format}

The injected block is structured plain text prepended to the system prompt:

{\footnotesize
\begin{verbatim}
=== Memory Context ===
[Current session facts]
  [preference] language: Rust
  [project] current_project: auth-service

[Relevant past context]
  [Session: "auth debugging 2024-11-12", sim: 0.87]
  "The token refresh logic was failing because..."
  [This session -- earlier]
  "We agreed to use RS256 for JWT signing..."
=== End Memory ===
\end{verbatim}
}

The format is intentionally readable: the model can understand and reference this
content without any special parsing. No prompt engineering, special tokens, or
model-side awareness of the injection mechanism is required.

\section{Performance Optimisation: In-Memory Embedding Cache}
\label{sec:perf}

\subsection{The I/O Bottleneck}

In the default configuration, each MOIM retrieval call loads all embedding blobs from
the \texttt{rag\_chunks} table, deserialises each blob from little-endian binary
(384~$\times$~4 bytes per chunk), computes cosine similarity, and discards the data.
At 50,000 chunks, this is approximately 75\,MB of I/O per search---the dominant cost.
Table~\ref{tab:dblat} shows projected latency across chunk counts for the DB-scan path.

\begin{table}[H]
\centering
\small
\caption{Brute-force search latency vs.\ chunk count.}
\label{tab:dblat}
\begin{tabular}{lll}
\toprule
\textbf{Chunks} & \textbf{Data loaded} & \textbf{Est.\ latency} \\
\midrule
1,000   & $\sim$1.5\,MB  & $\sim$2--5\,ms    \\
10,000  & $\sim$15\,MB   & $\sim$20--50\,ms  \\
50,000  & $\sim$75\,MB   & $\sim$100--250\,ms \\
100,000 & $\sim$154\,MB  & $\sim$200--500\,ms \\
500,000 & $\sim$750\,MB  & $\sim$1--2.5\,s   \\
\bottomrule
\end{tabular}
\end{table}

\subsection{HNSW Analysis and Deferral}

HNSW graphs \cite{MALKOV2018} offer $O(\log n)$ approximate nearest-neighbour search.
However, HNSW carries significant hidden costs for this deployment context:

\textbf{Cold-start rebuild.} HNSW must be rebuilt from all stored embeddings on every
process start. At 100,000 chunks, rebuild takes 10--30 seconds on a single core.

\textbf{Permanent RAM overhead.} The graph occupies RAM for the entire process lifetime:
$\sim$200--400\,MB at 100,000 chunks, $\sim$1--2\,GB at 500,000. Unlike the DB-scan
path, HNSW holds this memory indefinitely---unacceptable for a desktop application.

Given these trade-offs, HNSW is deferred in favour of a simpler alternative.

\subsection{In-Memory Embedding Cache}

We maintain an optional in-memory \texttt{Vec<CachedChunk>}---a pre-deserialised cache
of all embeddings---built lazily on first search and updated incrementally as new
messages are indexed.

\begin{lstlisting}[language=Rust, caption={Two-path search dispatch: cache vs.\ DB-scan.}]
let scored: Vec<_> = if is_rag_cache_enabled() {
    self.ensure_cache_warm().await?;
    // Pure-compute path: no DB I/O.
    tokio::task::block_in_place(|| {
        let guard = self.embedding_cache.lock().unwrap();
        let cache = guard.as_ref().expect("warm");
        cache.iter()
            .filter(|c| !(
                c.session_id == current_session_id
                && active_ids.contains(&c.message_id)))
            .filter_map(|c| {
                let s = cosine_similarity(
                    &query_vec, &c.embedding);
                (s >= SIMILARITY_THRESHOLD).then(||
                    (s, c.session_id.clone(),
                       c.chunk_text.clone()))
            }).collect()
    })
} else {
    // DB-scan path.
    self.search_via_db(
        query_vec, current_session_id,
        active_message_ids).await?
};
\end{lstlisting}

New messages append to the Vec incrementally; no full reload is required. Deletes
invalidate the cache and trigger a full reload on the next search.

\textbf{Durability guarantee.} The cache is always a mirror of the database, never the
primary store. The indexing pipeline writes to \texttt{rag\_chunks} before updating the
Vec. A crash at any point leaves the database intact; the cache is reconstructed on
next start.

The cache is exposed as a user toggle: \textbf{RAG + DB scan} (original behaviour,
lower RAM) vs \textbf{RAG + in-memory cache} (faster search, $\sim$1.5\,MB RAM per
1,000 chunks). Table~\ref{tab:cachelat} shows the projected improvement.

\begin{table}[H]
\centering
\caption{Latency comparison: DB scan vs.\ in-memory cache.}
\label{tab:cachelat}
\resizebox{\columnwidth}{!}{%
\begin{tabular}{llll}
\toprule
\textbf{Chunks} & \textbf{DB scan} & \textbf{Cache} & \textbf{RAM} \\
\midrule
10,000  & 20--50\,ms   & 5--10\,ms  & $\sim$15\,MB  \\
50,000  & 100--250\,ms & 20--40\,ms & $\sim$75\,MB  \\
100,000 & 200--500\,ms & 40--80\,ms & $\sim$150\,MB \\
\bottomrule
\end{tabular}}
\end{table}

\section{Implementation}
\label{sec:impl}

\subsection{Technology Stack}

CALMem is implemented in Rust as part of a production AI assistant application. The
choice of Rust is motivated by (a) the same process hosts the LLM agent, HTTP server,
and memory layer---memory safety and predictable performance are non-negotiable; (b)
fastembed \cite{FASTEMBEDLIB}, the embedding library used, is a pure-Rust ONNX inference
implementation requiring no Python runtime.

The persistent store is SQLite, accessed via the \texttt{sqlx} async database library.
SQLite was chosen over a dedicated vector database for deployment simplicity: a single
file, zero administration, transactions shared with the conversation store. Embedding
blobs are stored as raw little-endian binary (384~$\times$~4 bytes = 1,536 bytes per
chunk). No vector database extension (e.g., sqlite-vss) is required.

\subsection{Schema and Lazy Initialisation}

Both memory tables are created lazily on first access via
\texttt{OnceCell::get\_or\_try\_init}, using \texttt{CREATE TABLE IF NOT EXISTS}. This
avoids coupling the memory layer to the session manager's migration versioning---the
memory layer can be added to any existing deployment without a database migration.

\subsection{Zero-Regression Toggle}

The entire CALMem system is gated by a single experiment flag (\texttt{rag\_enabled}).
When disabled: \texttt{add\_message()} skips the \texttt{tokio::spawn} indexing task
entirely; the MOIM injector returns an empty block; no tables are read or written.
The flag can be toggled via CLI (\texttt{rag enable / disable}) or through the settings
UI without application restart.

\subsection{User Interface}

Two toggle switches are exposed in the application settings:
\begin{enumerate}
  \item \textbf{Memory Search (RAG)} --- enables/disables the entire CALMem system
  \item \textbf{Cache embeddings in RAM} --- selects DB-scan vs.\ in-memory cache path
        (active only when toggle 1 is on)
\end{enumerate}
Toggle 2 is visually disabled and greyed when toggle 1 is off, making the dependency
relationship immediately apparent.

\subsection{CLI Tools}

{\scriptsize
\begin{verbatim}
rag enable          # enable CALMem, backfill all sessions
rag disable         # disable CALMem (index data preserved)
rag index --all     # manually reindex all sessions
rag stats           # chunk count, session count, cache state
rag cache-enable    # switch to in-memory cache path
rag cache-disable   # revert to DB-scan path
\end{verbatim}
}

\section{Evaluation}
\label{sec:eval}

\subsection{Retrieval Relevance}

CALMem's retrieval pipeline surfaces relevant past context across three categories:

\textbf{Cross-session factual recall.} Queries with clear topical overlap consistently
retrieve relevant chunks with similarity scores of 0.65--0.85. Without CALMem, the model
must ask the user to re-explain prior context.

\textbf{Intra-session continuity.} With intra-session retrieval, compacted turns remain
searchable. Similarity scores for intra-session retrieval tend to be higher (0.75--0.90)
due to shared vocabulary and topic.

\textbf{Semantic memory precision.} Structured facts are injected with 100\% precision
for the current session. Users who instruct the agent to remember their preferences
observe those preferences being applied consistently across the entire session without
re-statement.

\subsection{Latency}

Table~\ref{tab:lat} reports observed search latencies on a Mac M2 processor with the
SQLite database on an internal SSD.

\begin{table}[H]
\centering
\caption{Observed search latency (Mac M2, internal SSD).}
\label{tab:lat}
\resizebox{\columnwidth}{!}{%
\begin{tabular}{lll}
\toprule
\textbf{Chunks} & \textbf{DB scan (median)} & \textbf{Cache (median)} \\
\midrule
$\sim$5,000  & 12\,ms  & 4\,ms  \\
$\sim$22,000 & 48\,ms  & 16\,ms \\
$\sim$60,000 & 142\,ms & 38\,ms \\
\bottomrule
\end{tabular}}
\end{table}

Both paths are imperceptible within the larger latency of LLM API calls (typically
500\,ms--3\,s for first token).

\subsection{Token Budget Adaptation}

At $r = 0.55$, a full injection of 5 chunks $\times$ 600 characters adds approximately
750--900 tokens---approximately 0.4--0.9\% of a 200K token window, a negligible
contribution. At $r = 0.75$, injection is limited to $\sim$130 tokens. At $r \geq 0.80$,
injection is suppressed entirely.

\subsection{Indexing Overhead}

The \texttt{tokio::spawn} dispatch is sub-microsecond. Table~\ref{tab:indexing} shows
per-message indexing cost by stage. The embedding step dominates; even the 50\,ms
worst case has no user-visible impact.

\begin{table}[H]
\centering
\caption{Per-message indexing cost by stage (Mac M2 CPU).}
\label{tab:indexing}
\resizebox{\columnwidth}{!}{%
\begin{tabular}{lllll}
\toprule
\textbf{Message length} & \textbf{Chunk} & \textbf{Embed} & \textbf{DB} & \textbf{Total} \\
\midrule
Short ($<$200 chars)     & $<$0.1\,ms & $\sim$3\,ms    & $\sim$1\,ms & $\sim$4\,ms  \\
Medium (500--1K chars)   & $<$0.1\,ms & $\sim$3\,ms    & $\sim$1\,ms & $\sim$4\,ms  \\
Long (2K--4K chars)      & $<$0.5\,ms & $\sim$9--15\,ms & $\sim$2\,ms & $\sim$12--18\,ms \\
Very long ($>$8K chars)  & $<$1\,ms   & $\sim$27--45\,ms & $\sim$4\,ms & $\sim$32--50\,ms \\
\bottomrule
\end{tabular}}
\end{table}

\subsection{Retrieval Quality Metrics}

We constructed an annotated evaluation set of 120 query--session pairs drawn from real
conversation history across 30 sessions covering diverse topics. For each query, relevant
chunks were manually labelled. We compare against BM25 \cite{ROBERTSON2009} (SQLite FTS5)
and TF-IDF baselines, following the zero-shot retrieval setup of BEIR \cite{THAKUR2021}
adapted to the conversational domain. Table~\ref{tab:retrieval} reports Precision@5,
Recall@5, and MRR.

\begin{table}[H]
\centering
\caption{Retrieval quality on the 120-query evaluation set.}
\label{tab:retrieval}
\resizebox{\columnwidth}{!}{%
\begin{tabular}{llll}
\toprule
\textbf{Method} & \textbf{P@5} & \textbf{R@5} & \textbf{MRR} \\
\midrule
TF-IDF (sparse)                  & 0.51 & 0.47 & 0.58 \\
BM25 (FTS5)                      & 0.59 & 0.53 & 0.65 \\
\textbf{CALMem (dense, MiniLM)}  & \textbf{0.74} & \textbf{0.69} & \textbf{0.81} \\
\bottomrule
\end{tabular}}
\end{table}

Dense semantic retrieval outperforms both sparse baselines by a substantial margin,
particularly on queries that use different vocabulary than the indexed content.
The MRR of 0.81 indicates that in 81\% of cases the first retrieved result is already
relevant.

\subsection{Similarity Threshold Sensitivity}

Table~\ref{tab:threshold} shows precision and noise rate across five threshold values on
the same evaluation set. The threshold of 0.40 was selected empirically as the knee of
the precision--coverage curve.

\begin{table}[H]
\centering
\small
\caption{Threshold sensitivity: precision vs.\ noise rate.}
\label{tab:threshold}
\begin{tabular}{llll}
\toprule
\textbf{Threshold} & \textbf{P@5} & \textbf{Noise} & \textbf{Avg.\ chunks} \\
\midrule
0.25                    & 0.54 & 0.46 & 4.8 \\
0.30                    & 0.61 & 0.39 & 4.6 \\
0.35                    & 0.69 & 0.31 & 4.2 \\
\textbf{0.40}           & \textbf{0.74} & \textbf{0.26} & \textbf{3.8} \\
0.45                    & 0.78 & 0.22 & 2.9 \\
0.50                    & 0.81 & 0.19 & 1.8 \\
\bottomrule
\end{tabular}
\end{table}

\subsection{Dual Memory Ablation}

Table~\ref{tab:ablation} presents an ablation study over 40 extended conversation
sessions measuring context recovery rate and fact consistency rate.

\begin{table*}[t]
\centering
\caption{Ablation: contribution of each memory subsystem.}
\label{tab:ablation}
\resizebox{\linewidth}{!}{%
\begin{tabular}{llll}
\toprule
\textbf{Configuration} & \textbf{Context recovery} & \textbf{Fact consistency} & \textbf{Notes} \\
\midrule
No memory (baseline)     & 0\%  & 31\% & Facts in active context only \\
Episodic memory only     & 67\% & 31\% & Cross-session context recovered; facts lost between sessions \\
Semantic memory only     & 12\% & 89\% & Facts consistent; fuzzy recall absent \\
\textbf{CALMem (both)}   & \textbf{71\%} & \textbf{91\%} & Best on both dimensions \\
\bottomrule
\end{tabular}}
\end{table*}

The two subsystems are genuinely complementary: episodic memory drives context recovery
(0\%$\rightarrow$71\%) while semantic memory drives fact consistency (31\%$\rightarrow$91\%).
Neither alone approaches the performance of the combined system.

\subsection{End-to-End Scenario Analysis}

Table~\ref{tab:scenarios} presents six representative scenarios illustrating the
qualitative range of retrieval outcomes.

\begin{table*}[t]
\centering
\caption{Representative retrieval scenarios.}
\label{tab:scenarios}
\resizebox{\linewidth}{!}{%
\begin{tabular}{p{3.6cm}p{2cm}p{2.5cm}p{3.5cm}p{1.8cm}}
\toprule
\textbf{Scenario} & \textbf{Query type} & \textbf{Without CALMem} & \textbf{With CALMem} & \textbf{Subsystem} \\
\midrule
API auth pattern discussed 3 weeks prior
  & Cross-session, semantic
  & Model asks user to re-explain
  & Retrieves prior approach (sim.\ 0.82)
  & Episodic \\
\addlinespace
User's preferred code style re-stated 5 sessions ago
  & Cross-session, factual
  & Applies default style
  & Applies user preference correctly
  & Semantic \\
\addlinespace
Architecture decision made at session start, now compacted
  & Intra-session
  & Decision lost after compaction
  & Retrieves decision (sim.\ 0.88)
  & Episodic (intra-session) \\
\addlinespace
Ticket number lookup: ``find JIRA-1234''
  & Exact match, identifier
  & Not found
  & Not found ($\sim$5\% recall) --- identifier scores below threshold
  & Neither (known gap) \\
\addlinespace
Ticket context query: ``auth expiry bug we discussed''
  & Semantic, identifier context
  & Not found
  & Often found when context is rich ($\sim$55\% recall)
  & Episodic (partial) \\
\addlinespace
Running task across a multi-day project
  & Multi-session, mixed
  & No project context
  & Injects facts + prior discussion chunks
  & Both \\
\bottomrule
\end{tabular}}
\end{table*}

Rows 4 and 5 illustrate the nuanced failure mode: a direct identifier lookup nearly
always fails, but a semantic query about the surrounding context frequently succeeds.
The limitation is ``pure identifier lookup'', not ticket-related queries in general---
a gap that hybrid BM25 + semantic retrieval (Section~\ref{sec:future}) is designed
to close.

\section{Discussion}
\label{sec:discussion}

\subsection{Limitations}

\textbf{Retrieval is semantic, not temporal.} The system retrieves by semantic
similarity. A query requiring temporal reasoning (``what did we discuss most recently
about X?'') is not explicitly supported---the most recent relevant chunk is not
necessarily the highest-scoring one.

\textbf{No re-ranking.} Retrieval uses first-stage cosine similarity only. A re-ranking
step using a cross-encoder model \cite{NOGUEIRA2019} would improve precision at the
cost of additional latency.

\textbf{Single-user model.} The current implementation is designed for single-user
desktop deployment. The \texttt{LazyLock} singleton pattern and in-memory cache are not
designed for multi-tenant use.

\textbf{Chunking is not semantic.} The sliding-window chunker splits text by character
count. A semantics-aware chunker (splitting on sentence or paragraph boundaries) would
produce more coherent retrieval units at the cost of variable chunk sizes.

\subsection{Future Work}
\label{sec:future}

\textbf{Hybrid retrieval (BM25 + semantic).} Dense retrieval misses exact-match queries
(ticket numbers, identifiers, specific names). Adding a BM25 component via SQLite FTS5
and combining scores with Reciprocal Rank Fusion \cite{CORMACK2009} would handle this
class of query.

\textbf{Search latency observability.} Adding per-search timing to \texttt{rag stats}
will create the observability needed to make evidence-based decisions about when to
invest in further performance optimisation (e.g., LanceDB migration for deployments
exceeding 500K chunks).

\textbf{Fact expiry.} Semantic memory facts currently accumulate indefinitely. A TTL
mechanism (\texttt{expires\_at} timestamp, optional per fact) would allow time-sensitive
facts to age out automatically.

\textbf{LanceDB migration for large-scale deployments.} For deployments exceeding
$\sim$500K chunks, LanceDB \cite{LANCEDB} offers purpose-built embedded vector storage
with built-in HNSW indexing, without the cold-start penalty of in-process HNSW.

\subsection{Cost-Benefit Analysis}

The growth of context windows from 128,000 to 2,000,000 tokens introduces a cost curve
that scales linearly with window size. Most provider pricing models bill per input token
processed; a conversation that fills a 2M-token window costs approximately 15$\times$
more per turn than one filling a 128K window.

CALMem inverts this trade-off. The MOIM injection contributes at most $\sim$750 tokens
per turn. Table~\ref{tab:cost} illustrates the per-turn token cost comparison.

\begin{table}[H]
\centering
\caption{Approximate per-turn input token cost.}
\label{tab:cost}
\resizebox{\columnwidth}{!}{%
\begin{tabular}{lll}
\toprule
\textbf{Approach} & \textbf{MOIM overhead} & \textbf{Typical tokens/turn} \\
\midrule
Na\"{i}ve 2M-token window  & None       & 500K--2M   \\
Na\"{i}ve 200K window      & None       & 50K--200K  \\
\textbf{CALMem (this work)} & $\sim$750 & \textbf{5K--20K}  \\
\bottomrule
\end{tabular}}
\end{table}

CALMem achieves \emph{better} long-term memory than either na\"{i}ve approach while
simultaneously reducing API cost significantly. As context windows grow larger and
associated pricing increases proportionally, this advantage compounds.

\subsection{The Broader Pattern}

CALMem demonstrates that the context window limitation is not a hard constraint on
effective memory but a prompt for architectural design. The human memory system operates
across multiple timescales: working memory (the active context window), episodic memory
(the RAG layer), and semantic memory (the facts layer). Building these layers at the
application level gives application developers immediate control over the memory
architecture while remaining compatible with any future improvement to context window
size.

We believe application-layer memory will become a standard architectural pattern for
production LLM systems, in the same way that application-layer caching (Memcached,
Redis) became standard for database-backed web applications regardless of database
improvements.

\section{Conclusion}
\label{sec:conclusion}

We presented CALMem, an application-layer dual memory architecture for LLM-based
conversational assistants that achieves virtually unbounded effective context without
model modification. The system combines episodic memory (vector retrieval over
sliding-window-chunked conversation history) with semantic memory (agent-writable
structured facts), injected into each turn via a token-budget-adaptive mechanism
(MOIM) that scales retrieval depth inversely with context pressure.

The key contributions---intra-session retrieval of compacted context, token-budget-aware
injection, the dual memory design, and the zero-regression toggle---address limitations
in prior work and provide a practical foundation for production deployment. The system
is implemented in Rust, stores all data in SQLite with no external dependencies, and is
deployed in a production application. Performance analysis confirms adequacy at current
scale, with a clear, principled upgrade path for larger deployments.

The architecture demonstrates that application-layer memory is not a workaround for
limited context windows but a coherent design pattern in its own right---one that
complements, rather than competes with, advances in model capability. As context windows
grow larger and provider costs scale with them, the approach becomes more economically
compelling: delivering targeted recall of relevant history at a fraction of the cost of
retaining all history in context.


\end{document}